\documentclass[12pt]{article}   

\usepackage[T1]{fontenc}
\usepackage[utf8]{inputenc}
\usepackage{lmodern}
\usepackage{graphicx}
\usepackage{amsmath,amssymb,mathtools,bm}
\usepackage{siunitx}
\usepackage[numbers,sort&compress]{natbib}
\usepackage[hidelinks]{hyperref}               
\usepackage[nameinlink,capitalise,noabbrev]{cleveref}
\usepackage{booktabs,tabularx,array}
\usepackage{microtype}
\usepackage{xcolor}
\usepackage{url}

\providecommand{\bmsection}[1]{\section*{#1}}
\newenvironment{backmatter}{}{}

\graphicspath{{figs/}{figures/}{images/}}

\usepackage[numbers]{natbib}

\providecommand{\journalname}[1]{}

\begin{document}
\pagenumbering{arabic}

\title{Orthogonal Photoelastic Imaging for Three-Dimensional Stress Estimation in a Transparent Cubical Block}

\author{Dhiraj K.~Singh\thanks{Corresponding author: \href{mailto:u6021818@utah.edu}{u6021818@utah.edu}}\\
Department of Mechanical Engineering, University of Utah,\\
Salt Lake City, UT 84112, USA}

\date{} 
\maketitle
\thispagestyle{plain}  

\begin{abstract}
Conventional photoelastic methods are largely limited to two-dimensional stress visualization, leaving a gap in techniques that can capture three-dimensional force interactions with high sensitivity at low stress levels-capability that is critical for biomechanics and dynamic force analysis. This study develops and demonstrates a cubic photoelastic model that enables accurate fringe-order estimation from three orthogonal views, providing a foundation for reconstructing full three-dimensional stress states. A transparent, low-elasticity epoxy cube, free of prestress, was fabricated and examined using combined transmission and reflection photoelastic imaging. Three mutually orthogonal isochromatic fringe fields were recorded simultaneously under a single applied load. Image analysis employed a peak–valley intensity method to extract sub-fringe orders and to resolve low-stress cases with minimal noise. The cubic block produced high-quality fringe patterns in all directions, enabling separation of tangential and normal stress components. Independent orthogonal views confirmed directional sensitivity and yielded consistent fringe-order estimates under low loading, with response times on the order of tens of microseconds. These results establish a practical approach for three-dimensional photoelastic stress measurement from orthogonal views and create a pathway toward full vector force reconstruction with strong potential for biomedical applications and studies of dynamic loading.
\end{abstract}

\enlargethispage{-2\baselineskip}

\section[Introduction]{Introduction}

Digital photoelasticity is a widely adopted optical technique for experimental stress analysis in transparent materials subjected to mechanical loading~\citet{Ramesh2020}. It builds upon classical photoelasticity by leveraging digital image acquisition and computational processing to extract quantitative stress fields. Stress information is encoded in fringe patterns-\emph{isochromatics}, which represent the magnitude of principal stress differences, and \emph{isoclinics}, which reveal principal stress orientations~\citet{Scafidi2015}.

Over the years, several methods have been developed to demodulate fringe patterns, including fringe counting~\citet{Idichandy1983}, load stepping techniques~\citet{Ekman1998}, color-based twelve-fringe photoelasticity~\citet{Ramesh2015}, regularized phase tracking~\citet{Tian2014}, computational hybrid strategies~\citet{Restrepo2021}, and more recently, deep learning algorithms~\citet{Brinez2022}. These advances have made digital photoelasticity a robust and accurate tool for stress measurement across both static and dynamic contexts.

Applications of digital photoelasticity span a wide range of fields, including stress analysis in 3D-printed models~\citet{Ju2023}, geological rock evaluation~\citet{Ju2019}, tensile and compression testing~\citep{Ren2021, Kim2021, Amini2022}, and industrial diagnostics~\citep{Patterson2022,Wang2022,Ramesh2015}. Despite these advances, the use of photoelasticity for \emph{contact force measurement}-particularly with high spatial and temporal resolution-remains underexplored.

Contact force measurement is critical in many disciplines, such as biomechanics, robotics, sports science, and transportation. Traditional sensors (e.g., strain gauges, capacitive load cells\citet{Cheng}) measure total force components, while pressure pads~\citet{Davis} and 2D photoelastic sheets~\citet{Ramesh} typically resolve only one component (usually normal force) of the surface traction. However, accurate control and analysis in grasping~\citep{Dollar, Prattichizzo, Hawkes, Hawkes2} or human neuromotor studies~\citep{Johansson, Srinivasan} require complete knowledge of normal and shear stresses. Similarly, assessing athlete performance and injury risk, or evaluating tire-road interactions in transportation, depends on spatially resolved, multi-component force measurements.

Classical photoelastic methods are capable of very high spatial and temporal resolution, limited primarily by material stiffness and camera resolution. However, most applications rely on a \emph{single optical view}, which only captures stress components normal to the direction of light propagation. This limitation introduces ambiguity, as different loading configurations can produce identical fringe patterns. Recent work has demonstrated normal and shear force reconstruction for point loads using 2D photoelastic sheets~\citet{Dubey}, but a full-field, multi-component reconstruction remains challenging.

To overcome these limitations, \emph{three-dimensional photoelasticity} has been proposed. By capturing fringe data from multiple optical paths, one can recover all three components of the surface traction vector. For example, Bandi et al.~\citet{bandi2017system} demonstrated force reconstruction using optimization techniques with multiple surfaces of a thick photoelastic block. However, practical implementation of 3D photoelastic sensing has been hindered by a lack of standardized methods for \emph{fabricating, characterizing, and testing photoelastic blocks} with desirable optical and mechanical properties.

In this study, we present a fabrication method for transparent photoelastic blocks of arbitrary dimensions with a controlled elastic modulus and minimal optical distortion. This platform enables the acquisition of three independent isochromatic fringe fields associated with the normal and shear stress components. For each prescribed load applied to the cube specimen, the fringe order and the corresponding stresses are quantified across all three camera views, providing consistent cross-view evaluation. The resulting sensing system offers high sensitivity, rapid temporal response, and high spatial resolution-enabling accurate, multi-component contact force measurements in dynamic environments.

\subsection[Background \& Motivation]{Background \& Motivation}
\begin{figure*}[!ht]
    \centering
       \includegraphics[scale=1.6]{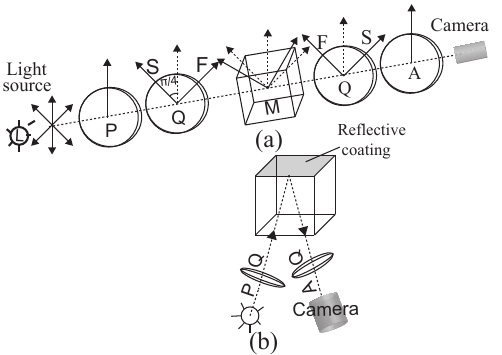}
        \caption{ (a) Schematic of bright circular polariscope:  polarizer and analyzer are parallel to each other, both quarter wave plates cross to each other, and the angle between polarizer and quarter wave plate is $\pi$/4. (b) Schematic of reflected polariscope: arrangement of polarizers and quarter wave plates is similar to bright circular polariscope but due to reflection from the top of the surface the angle changes and the bright field becomes a dark field. L - white diffuse light, P - polarizer, Q - quarter wave plate, M - photoelastic model, A - analyzer.}
\label{fig:polariscope}
\end{figure*}
Photoelasticity is a whole field method for estimating stresses and strains. Photoelastic materials are isotropic when free of stress but develop anisotropy under stress \citet{Ramesh}. The change in indices of refraction of the photoelastic material is linearly proportional to the applied load \citet{Maxwell}. When light falls on a photoelastic material it splits into two principal stress directions (perpendicular to each other) and each ray experiences a different refractive index. The two rays travel with different speeds and emerge with a phase difference that is a function of the wavelength of light and material properties \citet{Philips}. A color fringe pattern is obtained if white light is used to illuminate the stressed photoelastic material. 
\subsection[Related Work]{Related Work}
\subsubsection[Circular Polariscope Arrangements]{Circular Polariscope Arrangements}

An optical arrangement that exploits the properties of polarized light, called a polariscope, can be used to observe these fringe patterns. White diffuse light is used as an illumination source. We use two polarizers and two-quarter wave plates to create cross circular polarization. In our arrangement, the two polarizers are parallel to each other and the two-quarter wave plates are crossed with respect to each other to minimize error due to imperfect quarter wave plates. When an incident wave falls on the polarizer (P) with an angle $\alpha$, the polarized light beam that emerges from the linear polarizer can be represented by E=E$_0$cos($\alpha$).
Then, the light enters the quarter wave plate (Q), and is resolved into two components E$_f$ and E$_s$ with vibration parallel to the fast and slow axes. Since quarter wave plate is oriented  at $\pi$/4 to the polarizer axis, the light enters the first quarter wave plate and leaves with the two components having a phase difference $\pi$/2. After leaving the first quarter wave plate, the components of light vector enter the photoelastic model. Since the stressed block exhibits the characteristics of a temporary wave plate, the components of light again resolve into two. When it leaves the model an additional relative retardation ($\delta$) accumulates which is the difference of principal stresses \citep{Ramesh,Philips}. The light emerging from the model propagates to the second quarter wave plate (Q). As mentioned earlier, the second quarter wave plate is crossed with respect to first quarter wave plate. Therefore, it has an effect of de-rotating the light that was rotated by the first quarter wave plate. So retardation developed by first quarter wave plate is canceled by the second quarter wave plate. After the second quarter wave plate, light wave has retardation only due to stressed model. Then the component of the light wave parallel to the analyzer axis emerges from the analyzer, with intensity given by I=E$_0$$^2$ cos$^2$($\delta$/2). The intensity of the light beam emerging from the circular polarizer is a function of only principal stress difference and is independent of the angle between the analyzer and the direction of principal stress. Therefore, in the absence of model or for a model that is unloaded, the observed intensity is maximum. This configuration is called bright field circular polarization. This arrangement is used for transmitted intensity (camera1 and camera 2) and a schematic of this arrangement is shown in Fig.~\ref{fig:polariscope}a. A reflected polariscope is also used to extract optical signal from camera 3.  In the reflected polariscope, the polarizer and quarter wave plate are set as above. But, due to reflection of light from the top surface of the photoelastic block, the phase angle changes and leads to dark-field circular polarization. This is shown in Fig.~\ref{fig:polariscope}b. For this study, three Nikon D800E cameras, each with a resolution of 7360 $\times$ 4912 pixels, were used to capture three-dimensional optical fringe images. All cameras were equipped with a fixed focal length Nikon AF-S VR Micro-Nikkor 105 mm f/2.8G IF-ED lens.

\subsection[Contributions \& Paper Organization]{Contributions \& Paper Organization}
This paper introduces a cubic photoelastic framework for estimating three-dimensional forces using fringe order analysis under low-stress conditions. We demonstrate a peak–valley intensity method for accurate fringe order estimation, supported by calibration of elastic modulus, Poisson’s ratio, and photoelastic constant. The paper is organized as follows: Section~2 reviews related literature; Section~3 describes materials and methods; Section~4 characterizes the photoelastic block; Section~5 details the experimental setup; Section~6 presents fringe order and stress estimation results; and Section~7 concludes with key findings.

\section[Materials \& Methods]{Materials \& Methods}
\subsection[Block Fabrication]{Block Fabrication: Transparent Photoelastic Specimens}
\begin{figure*}[!ht]
    \centering
       \includegraphics[scale=0.5]{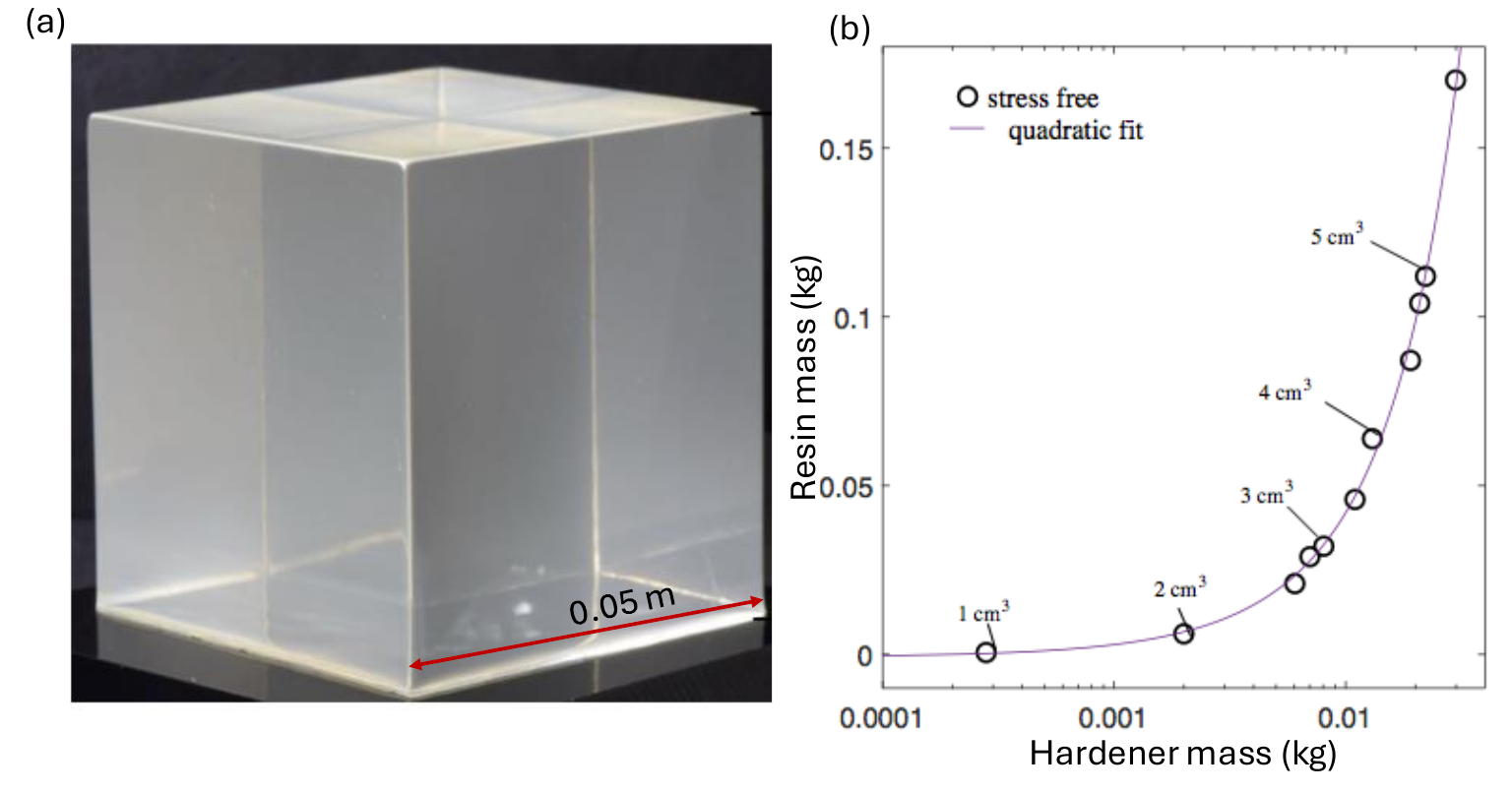}
 \caption{(a) Image of photoelastic cubic block with size 0.05 m made by resin/hardener solution with all required properties including, free of internal-stress, good transmittance, smooth sides, sharp edges, and no bubbles. (b) Hardener mass and resin mass plotted for cubic block of size 0.01 m, 0.02 m, 0.03 m, 0.04 m and 0.05 m (open circle), and the solid line is quadratic fit of hardener mass and resin mass.}
   \label{fig:method}
\end{figure*}
2-D photoelastic sheet can be easily manufactured. Manufacturers of resin/hardener have a standard protocol to make thin photoelastic sheet.  In a 2-D photoelastic sheet, only one direction (normal to the plane) is transparent and stress-free; the other directions (along with the plane) have poor transparency and nonzero internal stresses. Photoelasticity requires transparency and zero internal stresses in all three directions.  First, we used  PL-1 and PL-2 liquid plastic from Vishay Precision Group \cite{Vishay} but these polymers became opaque and developed pre-stress when the thickness of model increased beyond approximately 0.01 m. Preliminary experiments found that the volume ratio of resin/hardener, baking temperature and material of the mould are important aspects in making a photoelastic block with good transparency and in zero internal stress.
The photoelastic block was constructed as follows. {\bf Apparatus and materials}: mould, measuring jar, beaker, U-spacer, stirrer, resin/hardener and cleaner. {\bf Mould}: The photoelastic block becomes opaque when it comes in contact with rough side walls of the mould. So the selection of mould is also an important parameter. The material of mould was chosen as silicon, which has low thermal conductivity, low chemical reactivity, high thermal stability, non-sticky and smooth walls (Shin-Etsu silicone). The outer frame of silicon mould is 3D printed (Stratasys Dimension 1200es) with ABSPlus (Acrylonitrile butadiene styrene) thermoplastic material. The mould supports the photoelastic model as it sets with flat sides. {\bf Resin and hardener}: We used resin epoxy SQ-2001 and hardener SQ-3154 (manufacturer by Redelease ) to build a final photoelastic block. {\bf Ratio of resin/hardener}: The curing process of this epoxy resin/hardener is an exothermic reaction \citet{Corci}) and  produces heat that needs to be controlled to avoid thermal degradation. The amount of heat generated depends on the ratio of resin/hardener. Too little hardener provides insufficient energy to cure the photoelastic block and too much hardener provides extra energy that may break the bond between molecules and give permanent internal stress. So the correct ratio of resin/ hardener is required to build photoelastic blocks free of internal stress. We found, through trial and error, the correct ratio of resin to hardener to achieve zero pre-stress that depends on the size of the blocks. {\bf Procedure}: All components (mould, beakers, stirrer, funnel) were washed in propanol followed by methanol three times and dried in an oven for 10 minutes at 75$^{\circ}$C. Amount of resin and hardener were estimated for the corresponding size of the block. The mould with the outer frame was kept inside the oven at 75$^{\circ}$C. The epoxy resin and hardener were taken in two different polypropylene disposable beakers and kept inside an oven for 30 minutes at 75$^{\circ}$C to remove moisture.  The hardener was poured carefully avoiding introduction of air bubbles into the resin and the solution was stirred slowly for 5 minutes or until homogeneous. After mixing, a solution of the resin/hardener was kept aside for degassing for 60 seconds then the solution was poured into the mould using the funnel at a slow and steady rate.  Approximately after 5 hours photoelastic block was taken out (by using compressed air)  and  desiccated to avoid pre-stressed due to humidity. follow the procedure mention above, a zero internal-stress, transparent and smooth photoelastic block is obtained Fig.~\ref{fig:method}a. The relationship between the mass of epoxy resin and hardener required is plotted in Fig.~\ref{fig:method}b.

\subsection{Fringe Order Estimation Using Peak and Valley Intensities}

Fringe patterns observed in photoelastic experiments are the result of light interacting with stress-induced birefringence in transparent materials. The resulting intensity variation along a light path can be analyzed to determine the fringe order, which is proportional to the difference in principal stresses. For materials under low stress, where the fringe order is less than one, precise estimation becomes challenging due to the absence of distinct fringes. To address this, we adopt a peak-and-valley intensity method to quantify fractional fringe orders.

For each horizontal pixel trajectory (intensity profile), the intensity data is first smoothed using the \texttt{smoothdata} function in MATLAB to suppress noise while retaining the underlying fringe structure. Local peaks and valleys are then detected using \texttt{findpeaks} with tuned parameters for prominence and peak separation to suit low-order regions.

If the profile contains at least one valley, the fringe order $N$ at the point of interest is computed as the ratio of the relative intensity amplitude to a reference amplitude:
\begin{equation}
N = \frac{I_\ell - I_{v+1}}{I_f - I_v},
\end{equation}
where $I_\ell$ is the intensity at the point of interest, $I_{v+1}$ is the adjacent valley intensity, and $I_f$ and $I_v$ represent the intensities at a nearby fringe peak and valley, respectively.

In cases where no valley is detected along the pixel trajectory, typically for fringe orders below 0.5, a global minimum intensity $I_{\text{min}}$ from the full image sequence is used in place of $I_v$. The fringe order is then estimated as:
\begin{equation}\label{eq:subfringe}
N = \frac{I_\ell - I_{v+1}}{I_f - I_{\text{min}}}.
\end{equation}

This approach enables accurate estimation of sub-fringe order levels in low-stress zones, allowing the resolution of stress gradients even in materials subjected to mild loading.

 \section{Characterization}
\subsection{Internal Stress}
\begin{figure}[h!]
    \centering
       \includegraphics[scale=1.6]{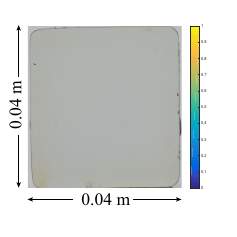}
 \caption{ Photoelastic cubic of size 0.04m block visualize through bright circular polariscope without loading and showing free of internal stress with approximately constant intensity entire the block.}
 \label{fig:prestress}
\end{figure}
To test the internal-stress in the photoelastic block, a cube block of size 0.04 m was illuminated under a bright circular polariscope without loading it. The circularly polarized light entered from one side of block and after crossing the entire block, an image was taken from the opposite side. Image looks bright with approximately uniform intensity that indicate no internal stress in the photoelastic block. The resultant images are shown in Fig.~\ref{fig:prestress} which does not indicate variation in intensity and hence no internal stress.
\subsection{Calibration: Modulus of elasticity, Poisson ratio, and photoelastic material constant}
\begin{figure}[h!]
    \centering
       \includegraphics[scale=0.5]{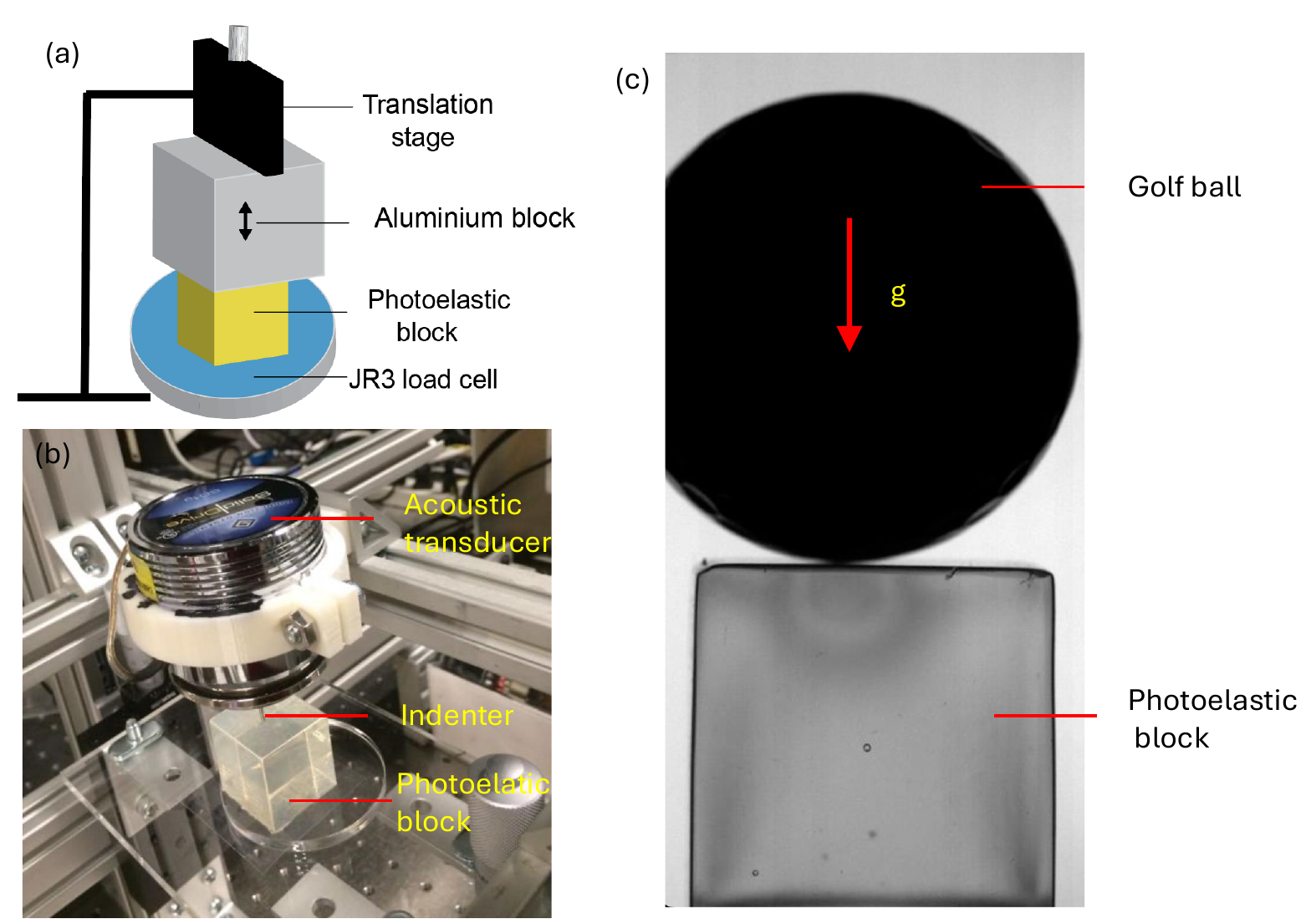}
 \caption{(a) Schematic of the experimental setup for calibration of the modulus of elasticity and Poisson ratio. (b) Experimental setup used to measure the response time of the photoelastic block. Acoustic transducer attaches with indenter used to give a high-frequency pulse on the photoelastic block and the optical signal are visualize through high-speed camera. Photoelastic block kept on plate and plate is placed on two translation stages to maintain level and proper contact between indenter and photoelastic block. (c) Golf ball impact on the photoelastic block, used to measure response time.}
 \label{fig:calib}
\end{figure}

\begin{figure}[h!]
    \centering
       \includegraphics[scale=1.6]{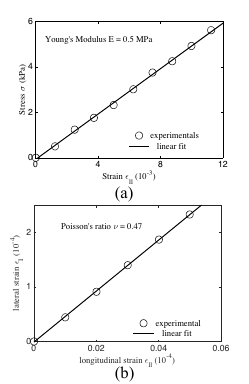}
 \caption{ Calibration curves for (a) elastic modulus: stress $\sigma$ (N/m$^2$) versus compressional or longitudinal strain $\epsilon_{||}$ (dimensionless) and (b) Poisson ratio: Lateral strain $\epsilon_{\perp}$ (dimensionless) vs compressional or longitudinal strain $\epsilon_{||}$ (dimensionless).}
 \label{fig:elasticity}
\end{figure}

A cubic block on the 0.04 m side was used for calibration. The block was kept on a JR3 load cell with metal base support. A schematic of this experimental setup is shown in Fig.~\ref{fig:calib}a. The compression method was used to calibrate the modulus of elasticity. For uniform distribution of force over the photoelastic block, a aluminium (AL) cube block bigger than the photoelastic block was used to compress it. Note that compression in the AL block was negligible and examined without a photoelastic block with a similar range of force. The AL block was attached to a vertical translation stage to provide a known longitudinal displacement.  In each step, the block was compressed by 0.00005 m and resultant force was measured. The force-displacement curve is plotted in Fig.~\ref{fig:elasticity}a. A linear fit gives the modulus of elasticity.
For Poisson ratio calibration, the same specimen was used and experimental setup was further modified by attaching a camera (D800E) with a macro lens and millimeter scale across the specimen to monitor the lateral strain perpendicular to the applied strain (longitudinal strain). A  compressive longitudinal strain was applied in steps of 0.00005 m, and the corresponding lateral strain was obtained. The resultant lateral strain and longitudinal stain is plotted in Fig.~\ref{fig:elasticity}b. A linear fit gives a Poisson ratio (nu = 0.47).  \newline
Calibration of photoelastic material of fringe constant (f$_{\sigma}$):
For the photoelastic material constant, a disc of 0.04 m diameter and 0.01m thickness is made keeping the same ratio of resin/hardener of corresponding to cubic block.
We have, from the stress optics law, 
\begin{equation}\label{eq3}
f_{\sigma}  = h(\sigma_{1} - \sigma_{2})/N       
\end{equation}
where $\sigma_{1}$ - $\sigma_{2}$ is the difference between principal stresses in the plane normal to the light propagation direction, h is the thickness of the disc and N is the order of fringes at the center of the disc.
Isochromatic fringe pattern with bright polariscope configuration is considered under diametrical compression (Conventional method) \cite{Murthy}. giving
\begin{equation}\label{eq4}
{\sigma_{1}} - {\sigma_{2}}  =  8P/{\pi}Dh 
\end{equation}
where, D is the diameter of the disc and P is the applied load on the disc. from Eq. (\ref{eq3})  and Eq. (\ref{eq4})

\begin{equation}\label{eq5}
f_{\sigma} = 63.69 \, \frac{P}{N} \;\; \text{N\,m}^{-1}\,\text{fringe}^{-1}
\end{equation}

 \begin{figure}[h!]
    \centering
       \includegraphics[scale=1.6]{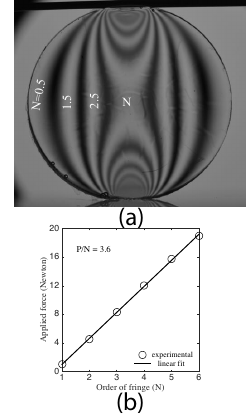}
 \caption{Calibration of photoelastic material constant (a) Image of fringe pattern (in green channel) of disc under diametrical compression. (b) order of fringe at the centre of disc versus load.}
 \label{fig:fringe}
\end{figure}
  
To determine the photoelastic fringe constant ($f_{\sigma}$) and evaluate the sensitivity and linearity of the photoelastic block, we applied a series of known loads to a circular disk and captured the resulting fringe patterns. The fringe order ($N$) at the center of the disk was estimated for each applied load ($P$) using an intensity-based method, as shown in Fig.~\ref{fig:fringe}a. As the load on the disk was gradually increased, the corresponding intensity at the center was recorded.

By plotting the fringe order against the applied load (Fig.~\ref{fig:fringe}b), we obtained a linear relationship, confirming the expected stress-optic behavior. The slope of the fitted line was found to be 3.6 fringe/N, which, when substituted into Eq. (\ref{eq5}), yields a photoelastic fringe constant of $f_{\sigma} = 229.3$ $\text{N\,m}^{-1}\,\text{fringe}^{-1}$. The fringe constant characterizes the material's sensitivity to stress: a lower value of $f_{\sigma}$ indicates higher sensitivity. Thus, the measured value confirms that the fabricated photoelastic block is suitable for detecting low-magnitude forces with high accuracy.

\subsection{Transparency, sensitivity, resolution and response time}

 The transmittance is estimated by measuring the intensity of light under similar conditions without and with the photoelastic block. For a cube block of size  0.04 m, we find that the light intensities are 190  $\pm$ 6 and 160 $\pm$ 6 without and with photoelastic block respectively.  To account for multiple reflections and transmissions, we write\begin{equation}\label{eq6}
160 = 190 ({\alpha} +  {\alpha} (1 - {\alpha}){^2} + ................)      
 \end{equation}
where $\alpha$ is the transmittance of the photoelastic block. The transmissivity of this block is 0.81 for white light. \newline
\subsection{Response Time}
For dynamic applications, the effective response time of the photoelastic model is a critical parameter. Two time scales govern what is observed in images: (i) the intrinsic opto–mechanical delay of the photoelastic material before a measurable change in birefringence appears, and (ii) the time required for the applied load to exceed the visibility threshold (i.e., the smallest stress change that produces a detectable fringe/intensity change above background noise). To isolate the intrinsic delay, we designed experiments that minimize the threshold-crossing time by using impulsive loading and high frame-rate imaging.

\textbf{Impact test.} A hard spherical impactor (golf ball) was dropped onto the photoelastic cube (Fig.~\ref{fig:calib}c) from heights \(h=1\text{–}4~\text{m}\) (impact velocities \(\approx\sqrt{2gh}\)), while a high-speed camera recorded the optical field at frame rates between \(40{,}000\) and \(80{,}000~\mathrm{fps}\) (representative data at \(50{,}000~\mathrm{fps}\)). The onset of mechanical contact was defined by the first frame showing geometric contact of the ball with the surface, and the optical onset was defined as the first frame in which the fringe intensity exceeded the pre-impact baseline. Across all heights and frame rates, the measured delay between mechanical contact and optical response was approximately \(20~\mu\text{s}\). At \(80{,}000~\mathrm{fps}\) (frame period \(12.5~\mu\text{s}\)), this delay corresponds to roughly one–two frames, indicating that the intrinsic response is on the order of tens of microseconds and near the temporal resolution of our imaging.

\textbf{Acoustic validation.} As a complementary test with a controlled, repeatable input, an acoustic transducer was used to excite the block at \(2\text{–}4~\mathrm{kHz}\) (Fig.~\ref{fig:calib})b). Image sequences acquired at \(40{,}000\text{–}80{,}000~\mathrm{fps}\) showed a consistent phase lag between the applied excitation and the optical signal that is compatible with a \(\sim 20~\mu\text{s}\) response, supporting the impact-test estimate.

\section{Experimental set-up}\label{sec:setup}

\begin{figure*}[!ht]
    \centering
       \includegraphics[scale=1.6]{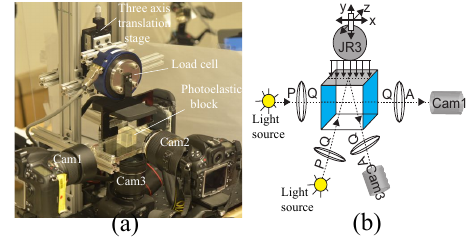}
 \caption{(a) Experimental setup used in extraction of three independent optical signals. Two transmitted camera 1(cam1) and camera 2 (cam2) and one reflected polariscope camera 3 (cam3) set across a photoelastic block. The block compress through knife-edge and knife-edge is attached with six-axis JR3 load cell and the load cell is attached to multi-axis translation stage to provide normal and shear forces. (b) Schematic of the experimental setup.}
 \label{fig:expsetup}
\end{figure*}

The photoelastic block was mounted in a metallic rig and clamped by an acrylic plate of dimensions $0.25\,\text{m} \times 0.25\,\text{m} \times 0.012\,\text{m}$. A six-axis load cell (JR3, Inc.) was connected to a three-axis translation stage to apply known normal and shear force distributions to the block surface. The opposite end of the load cell was fitted with a knife edge (Fig.~\ref{fig:expsetup}a). A schematic of the experimental setup is shown in Fig.~\ref{fig:expsetup}b. The knife edge was realized using a miniature precision shaft made of through-hardened 17-4 PH stainless steel for corrosion resistance and strength, with a $10\,\mu\text{in}$ RMS surface finish. Shafts of diameter $0.001\,\text{m}$, $0.002\,\text{m}$, and $0.004\,\text{m}$ (diameter tolerance: $-0.0005$ to $0\,\text{mm}$; straightness tolerance: $0.01\,\text{mm}/\text{m}$; McMASTER-CARR) were used.

Three Nikon D800E cameras, each equipped with a polarizer and a quarter-wave plate, and three diffuse white light sources, each with a polarizer and a quarter-wave plate, were arranged around the block to acquire three independent optical signals. Two transmission configurations and one reflection configuration were implemented to capture stress-fringe images during loading. The normal force was applied using the vertical translation stage; after verifying a nearly constant normal force, shear forces were introduced via the remaining two stages. Normal and shear forces were measured directly by the six-axis load cell.

All experiments were performed in a darkened laboratory to prevent entry of unpolarized ambient light. Here, the background intensity is defined as the light incident on the camera that does not pass through the block. To quantify it, an opaque plate of the same size as the block was placed in front of the setup and imaged using identical camera settings. Imperfections of the polarizer–analyzer pairs were also characterized: ideally, crossed polarizers yield zero transmitted intensity; any nonzero residual intensity observed in practice was measured and accounted for in the analysis.

\subsection{Application of Normal and Shear Loads}
Measurement of combined shear and normal force fields is essential in numerous applications; for example, predicting ulceration in diabetic patients requires knowledge of both shear and normal forces acting on the plantar surface of the foot~\citet{Schie}. While many point-load sensors can report forces along all three axes, they do not provide the full spatial stress field. Recent photoelastic approaches have estimated normal and shear forces for point loading using a single (normal) view~\citet{Dubey}; however, a single optical view is insufficient to fully resolve composite loading.

\begin{figure}[h!]
    \centering
       \includegraphics[scale=1.2]{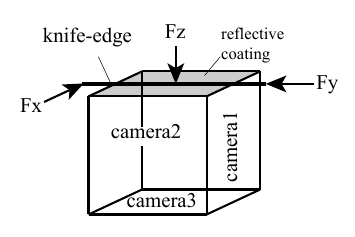}
 \caption{ Schematic of normal and shear force are provided on the photoelastic block.}
 \label{fig:sch}
\end{figure}

To address this, we employed the transparent photoelastic cubical block described above and acquired three independent optical signals to predict the surface shear-force field. The normal load \(F_z\) was applied quasi-statically using a vertical translation stage to establish a stable preload. Subsequently, in-plane shear loads \(F_x\) and \(F_y\) were introduced by incrementally translating two orthogonal horizontal stages. The schematic of applying all three loads to the photoelastic block is shown in Fig.~\ref{fig:sch}. In the multi-axis trials reported here, images were captured from Cameras~1–3 to document the fringe patterns; no quantitative fringe/stress inversion was performed for these cases.

\section{Results}

\subsection{Fringe Order and Stress Estimation Under Low Load}
 \begin{figure}[h!]
    \centering
       \includegraphics[scale=0.6]{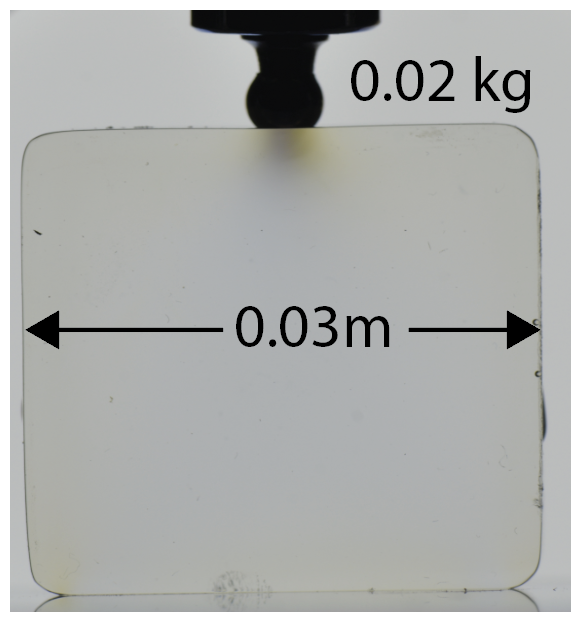}
 \caption{ Photoelastic cubic block with size 0.03 m loaded with 0.02 kg mass visualized through a bright circular polariscope.}
 \label{fig:resolution}
\end{figure}
\begin{figure}[htbp]
    \centering
    \includegraphics[width=\textwidth]{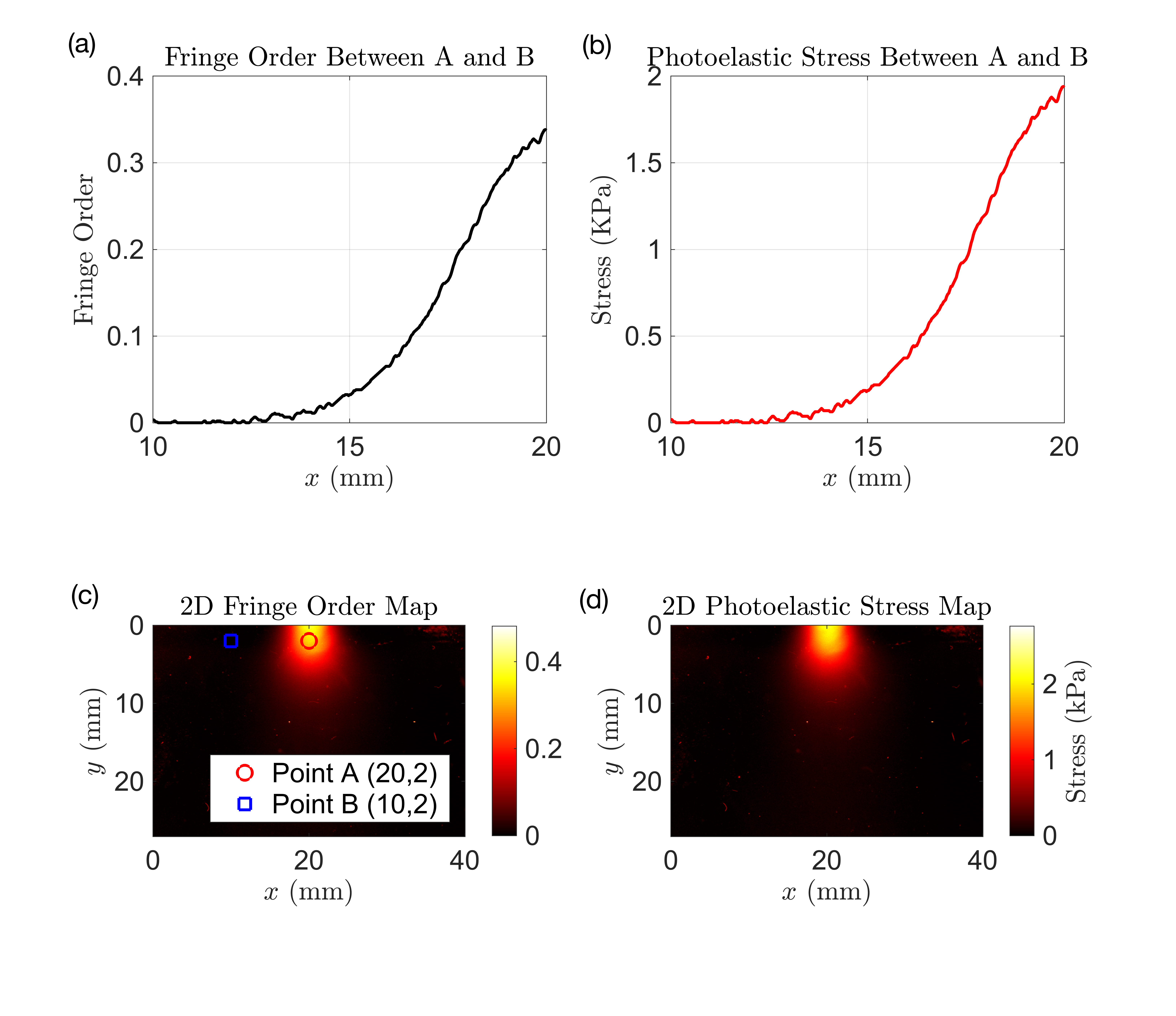}
    \caption{Fringe order and stress were estimated using the peak-and-valley intensity method. A 2-gram weight, shown in Figure~8, was applied at the center of the 4~cm~$\times$~4~cm photoelastic block. (a) Fractional fringe order between points A and B; (b) Stress distribution between points A and B; (c) Two-dimensional fringe order map; (d) Two-dimensional stress field.}
    \label{fig:fringe_order_stress}
\end{figure}

After successful manufacturing and calibration of the photoelastic block, the specimen was subjected to a low-magnitude point load, and the peak-and-valley intensity method was applied to estimate the fringe order and the corresponding stress field. The sensitivity of a photoelastic block refers to its ability to resolve small differences in force and to detect low levels of applied stress. To evaluate this sensitivity, a cubic block of size 0.04\,m was tested under an approximate load of 0.02\,N, as illustrated in Fig.~\ref{fig:resolution}.

In this low-stress scenario, the resulting fringe order was found to be less than 0.5. Accordingly, Eq. (~\ref{eq:subfringe}) was used to estimate the fringe order based on relative and reference amplitudes, and the stress field was computed using the known photoelastic constant. Two representative points, labeled 'A' and 'B', were selected on the fringe image to extract fringe order and corresponding stress values. The 1D fringe order profile, stress distribution, 2D fringe order map and 2D stress map are shown in Figs.~\ref{fig:fringe_order_stress}a, \ref{fig:fringe_order_stress}b, \ref{fig:fringe_order_stress}c and \ref{fig:fringe_order_stress}d, respectively.

\subsection{Fringe Order and Stress Analysis from Three Optical Views}

Mechanical testing of the photoelastic block was carried out under knife-edge unsymmetrical loading (line load distribution). Three independent optical images were acquired: two using transmitted polariscope arrangements and one using a reflected polariscope arrangement. Multiple viewing directions were required because stresses acting along the light propagation path cannot be visualized in the corresponding image. In each view, one force component is effectively hidden; therefore, a complete reconstruction of the contact forces necessitates three orthogonal views.  

\begin{figure}[htbp]
    \centering
    \includegraphics[width=\textwidth]{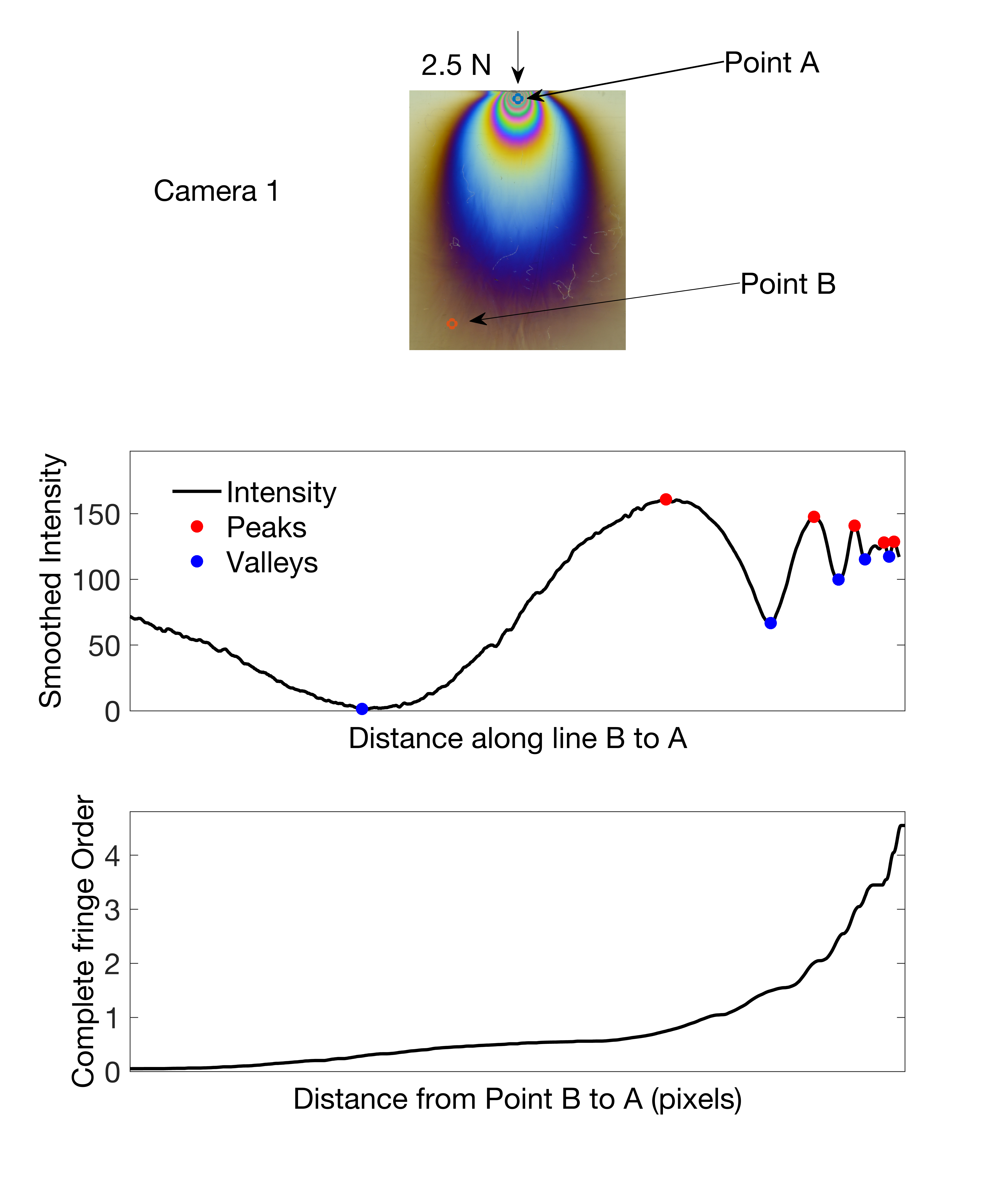}
    \caption{Fringe order estimation between two points under a 2.5~N applied load. 
    (Top) Photoelastic image captured by Camera 1 showing isochromatic fringe patterns, with Points A and B marked. 
    (Middle) Smoothed intensity profile along the line from Point B to A, with detected intensity peaks (red) and valleys (blue). 
    (Bottom) Estimated complete fringe order as a function of pixel distance from Point B to Point A, derived from peak-and-valley analysis.}
    \label{fig:fringe_order_estimation1}
\end{figure}

\begin{figure}[htbp]
    \centering
    \includegraphics[width=\textwidth]{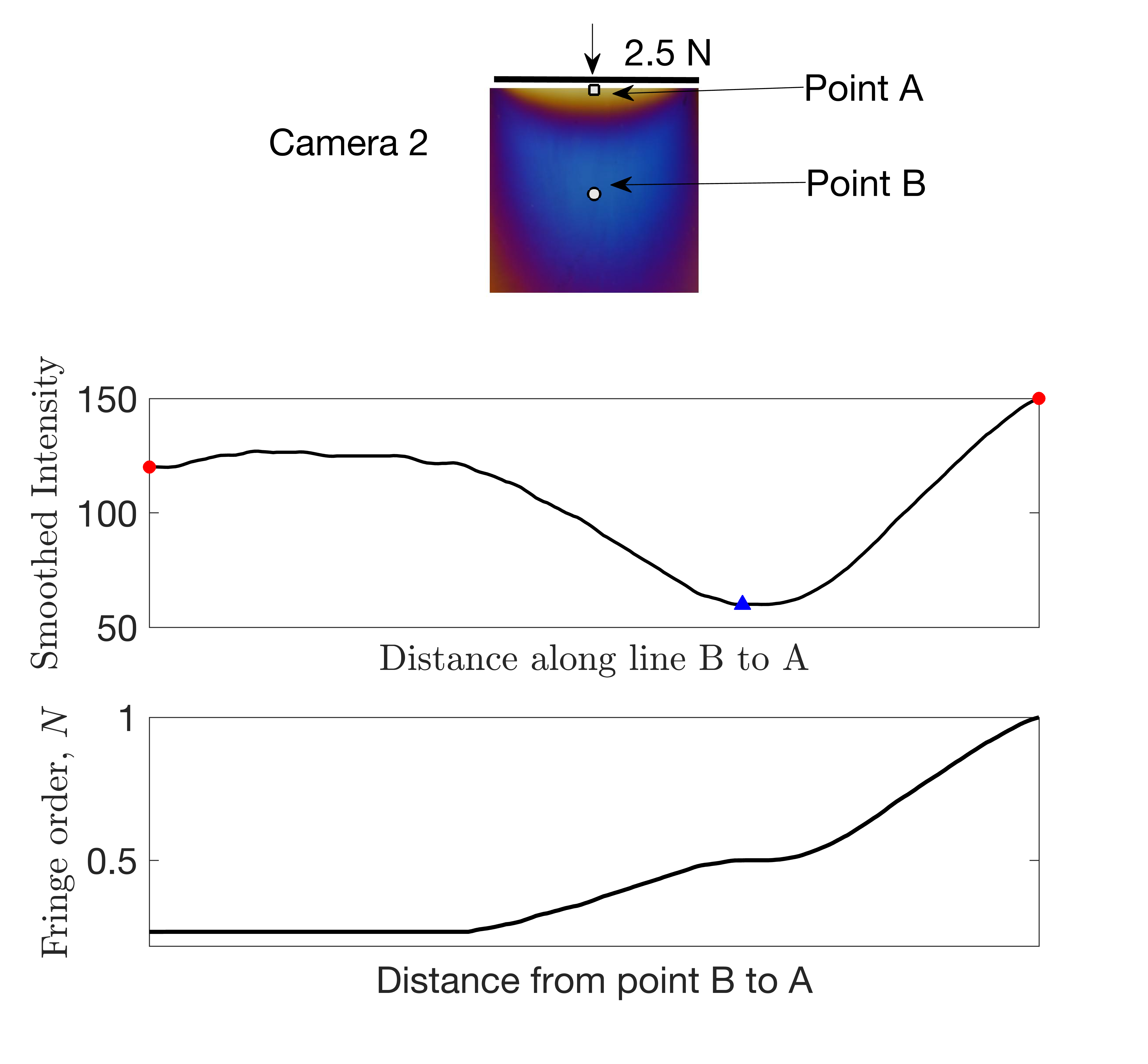}
    \caption{Fringe order estimation between two points under a 2.5~N applied load. 
    (Top) Photoelastic image captured by Camera 2 showing isochromatic fringe patterns, with Points A and B marked. 
    (Middle) Smoothed intensity profile along the line from Point B to A, with detected intensity peaks (red) and valleys (blue). 
    (Bottom) Estimated complete fringe order as a function of pixel distance from Point B to Point A, derived from peak-and-valley analysis.}
    \label{fig:fringe_order_estimation2}
\end{figure}

\begin{figure}[htbp]
    \centering
    \includegraphics[width=\textwidth]{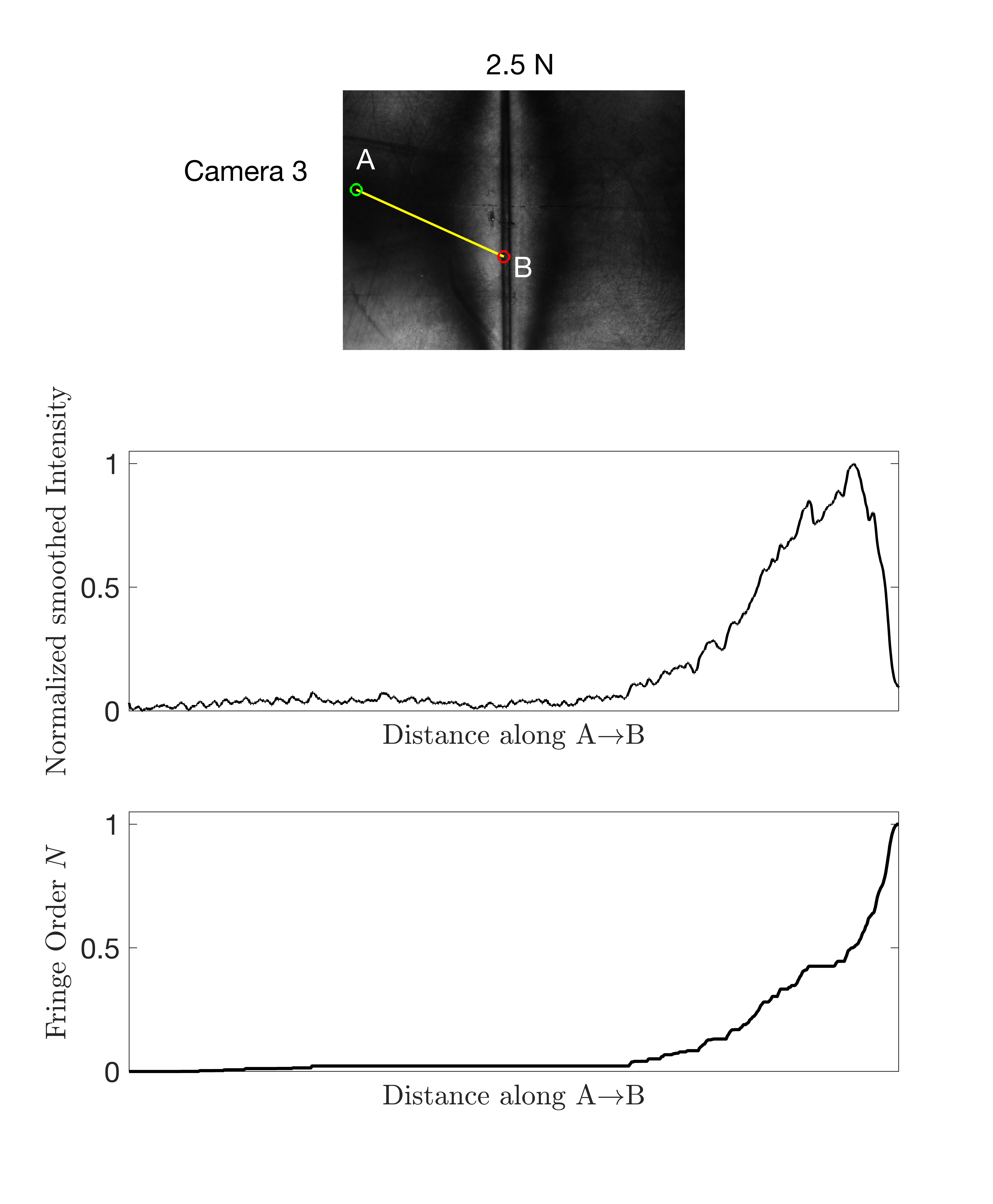}
    \caption{Fringe order estimation between two points under a 2.5~N applied load. 
    (Top) Photoelastic image captured by Camera 3 showing isochromatic fringe patterns, with Points A and B marked. 
    (Middle) Smoothed intensity profile along the line from Point B to A, with detected intensity peaks (red) and valleys (blue). 
    (Bottom) Estimated complete fringe order as a function of pixel distance from Point B to Point A, derived from peak-and-valley analysis.}
    \label{fig:fringe_order_estimation3}
\end{figure}
In the static configuration, a knife-edge load was applied (details provided in Section~\ref{sec:setup}). Although the magnitudes of the top and bottom loads were equal, the knife-edge geometry created asymmetric contact areas on the top and bottom surfaces of the block. This asymmetry produced distinct stress distributions and, consequently, different fringe patterns across the three camera views.  

\textbf{Camera 1 (Transmitted Polariscope):} The block experienced high stress on the top surface due to the small knife-edge contact area, and comparatively low stress on the bottom surface where the contact area was larger. The resulting fringe pattern appeared predominantly on the top side. Fringe order estimation was performed between two selected points along the stress gradient using the peak–valley intensity method. The smoothed intensity and reconstructed fringe order distribution are shown in Fig.~\ref{fig:fringe_order_estimation1}.  

\textbf{Camera 2 (Transmitted Polariscope):} In this view, both the top and bottom contact areas were equal, resulting in symmetric stress distribution. As a consequence, the observed fringe pattern was also symmetric. Fringe order analysis between two selected points was again carried out using the peak–valley method, with results presented in  Fig.~\ref{fig:fringe_order_estimation2}.  

\textbf{Camera 3 (Reflected Polariscope):} In the unloaded state, the reflected polariscope image appeared nearly dark, with intensity values close to zero. Upon application of load, the transmitted light intensity increased and clear fringe patterns emerged. Fringe order estimation followed the same peak–valley intensity method Eq. (~\ref{eq:subfringe}) as applied to Cameras 1 and 2. Figure.~\ref{fig:fringe_order_estimation3} shows the raw optical image, the corresponding smoothed intensity profile, and the final reconstructed fringe order distribution. These results confirm that the reflected configuration captures load-induced birefringence with sufficient clarity for quantitative stress analysis.  

\begin{figure*}[!ht]
    \centering
       \includegraphics[scale=1.2]{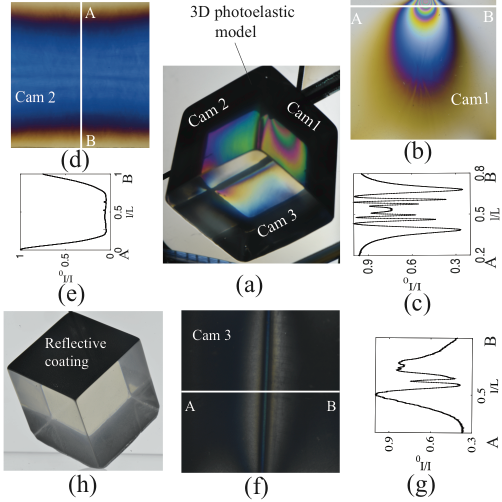}
       \caption{Three independent optical signals are obtained from the photoelastic block. (a) A three-dimensional image of stress fringe pattern through knife-edge loading.  (b), (d), (f) Images are captured after three independent polariscope arrangements across block for camera 1, camera 2 and camera3 respectively and in (c), (d) (f) corresponding normalized intensity plotted from A to B with standard deviation 0.04, 0.04 and 0.07 respectively. (h) Photoelastic block with the top surface reflective coating.}
 \label{fig:signal}
\end{figure*}

All three stress fringe images are presented in Fig.~\ref{fig:signal}b,d,f for Cameras~1–3, respectively. To further demonstrate data quality, intensity profiles were extracted along a line from Point~A to Point~B in the direction of maximum stress variation (Figs.~\ref{fig:signal}c,e,g). The standard deviation of intensity was 0.04, 0.04, and 0.07 for Cameras~1, 2, and 3, respectively. These low values indicate robust measurements despite potential fluctuations caused by light transmission through the 0.05~m thick photoelastic block, optical clarity variations, internal reflections from adjacent surfaces, and distortion. Overall, the combined three-view approach provides reliable estimation of fractional fringe orders and resolves the full three-dimensional loading configuration applied to the block.

\subsection{Shear force}
Under combined loading, each camera emphasized the shear component tangential to its viewing plane. In Camera~1, the fringe pattern was inclined in the direction of the \(x\)-shear, reflecting sensitivity to \(F_x\) in the presence of \(F_z\); the \(y\)-shear component \(F_y\) was not observed because it lies along the viewing normal of Camera~1. Conversely, Camera~2 showed fringes tilted with the \(y\)-shear, indicating sensitivity to \(F_y\) (with \(F_z\)), while being relatively insensitive to \(F_x\). Camera~3 provided the third independent optical signal. Because each of Cameras~1 and~2 captures only one in-plane shear component, combining the three independent views enables straightforward determination of the magnitude and net direction of the surface force vector field using photoelasticity. This multi-view approach has potential utility in biomedical contexts such as diabetic foot ulcer risk assessment, footwear design for people with disabilities, and reduction of injury and ulcer formation~\citep{Tang,Davis2}.

 \begin{figure*}[!ht]
    \centering
       \includegraphics[scale=1.4]{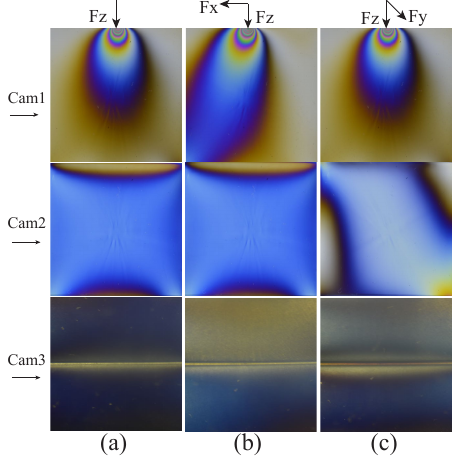}
 \caption{Thee independent optical view Images are captured with normal and shear forces (a) Only normal force, Fx = 0.02 N, Fy = 0.02 N, Fz = 3.85 N  . (b) Normal and shear force, Fx = 2.652 N, Fy = 0.03 N, Fz = 3.862 N. (c) Normal and shear force, Fx = 0.04 N, Fy = 2.469 N, Fz = 3.86 N.}
 \label{fig:shear}
\end{figure*}
\section{Conclusion}

We developed and validated a three–view, 3-D photoelastic platform and fabricated transparent blocks in multiple sizes with the required properties: high optical clarity, negligible optical/geometric distortion, minimal internal (residual) stress, and a low elastic modulus. From these specimens we acquired three independent optical signals with high spatial resolution. By selecting block size and thickness, the usable stress range and sensitivity can be tuned without compromising optical quality. Because the method responds to small strains with short temporal latency, the framework is well suited for dynamic \emph{fringe-order and stress} measurements.

In this study we did \emph{not} estimate forces. Instead, we estimated \emph{fringe order} and the corresponding \emph{stress} profiles between two points (A$\rightarrow$B) in all three camera views. Using the peak–valley intensity method (Sections~3.4 and~6.1–6.4) together with light, uniform smoothing, we achieved robust extremum detection and consistent sub-fringe interpolation via Eq.~(6). Fractional fringe order was resolved under a low externally applied load of \(0.2~\mathrm{N}\) (used only as a test condition, not inverted), demonstrating sensitivity in the low-strain regime relevant to thin photoelastic sensors intended to detect near-surface \emph{shear stress} in boundary-layer turbulence.

Across the three views we observed distinguishable optical patterns for no-load, predominantly normal loading, and predominantly shear loading (Sections~6.5–6.6). The no-load state produced near-uniform dark fields with zero fringe order in all cameras, while known normal/shear test cases produced different fringe evolutions and spatial distributions. These signatures enabled qualitative separation of load types and provide priors for future inverse estimation, but no force inversion was attempted here.

Material/optical characterizations (Sections~3 and~4) confirmed transparency and distortion targets, negligible internal stress, and calibrated Young’s modulus, Poisson’s ratio, and photoelastic constants—supporting quantitative stress mapping from measured fringe order. Looking ahead, the combination of three independent optical signals with calibrated material properties creates a clear path to future traction/force estimation via multi-view inverse solvers, along with uncertainty quantification for peak–valley detection and sub-fringe interpolation, and real-time implementation for dynamic environments. The present results establish that three-view photoelasticity, combined with careful fabrication and calibrated analysis, enables sensitive, high-resolution, and dynamically capable \emph{fringe-order and stress} measurements.

\begin{backmatter}

\bmsection{Acknowledgments}
The author thanks Kenneth Meacham III for exceptional experimental support. Experiments were conducted at the Collective Interactions Unit, Okinawa Institute of Science and Technology Graduate University (OIST), Japan, and the Department of Mechanical Engineering and Materials Science, Yale University, USA.

\bmsection{Supplementary material}
No supplementary material is associated with this work.

\end{backmatter}

\section*{Declarations}

\subsection*{Funding}
The authors received no specific funding for this work.

\subsection*{Conflicts of interest/Competing interests}
On behalf of all authors, the corresponding author states that there are no conflicts of interest.

\subsection*{Ethics approval}
Not applicable.

\subsection*{Consent to participate}
Not applicable.

\subsection*{Consent for publication}
All authors consent to publication.

\subsection*{Data availability}
Data and analysis code that support the findings of this study are available from the corresponding author upon reasonable request.

\subsection*{Author contributions}
All authors contributed to the study conception and design. Material preparation, data collection, and analysis were performed by the author team. All authors read and approved the final manuscript.


\bibliographystyle{unsrtnat}   

\bibliography{sample}






\end{document}